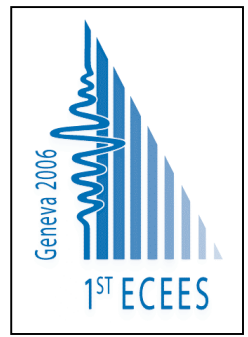

# IN SITU EXPERIMENT AND MODELLING OF RC-STRUCTURE USING AMBIENT VIBRATION AND TIMOSHENKO BEAM

Clotaire MICHEL[1], Stéphane HANS[2], Philippe GUEGUEN[3] and Claude BOUTIN[4]

### SUMMARY

Recently, several experiments were reported using ambient vibration surveys in buildings to estimate the modal parameters of buildings. Their modal properties are full of relevant information concerning its dynamic behaviour in its elastic domain. The main scope of this paper is to determine relevant, though simple, beam modelling whose validity could be easily checked with experimental data. In this study, we recorded ambient vibrations in 3 buildings in Grenoble selected because of their vertical structural homogeneity. First, a set of recordings was done using a 18 channels digital acquisition system (CityShark) connected to six 3C Lennartz 5s sensors. We used the Frequency Domain Decomposition (FDD) technique to extract the modal parameters of these buildings. Second, it is shown in the following that the experimental quasi-elastic behaviour of such structure can be reduced to the behaviour of a vertical continuous Timoshenko beam. A parametric study of this beam shows that a bijective relation exists between the beam parameters and its eigenfrequencies distribution. Consequently, the Timoshenko beam parameters can be estimated from the experimental sequence of eigenfrequencies. Having the beam parameters calibrated by the in situ data, the reliability of the modelling is checked by complementary comparisons. For this purpose, the mode shapes and eigenfrequencies of higher modes are calculated and compared to the experimental data. A good agreement is also obtained. In addition, the beam model integrates in a very synthetic way the essential parameters of the dynamic behaviour.

### 1. CONTINUOUS MODELLING OF BUILDINGS

The in situ monitoring is an easy way for collecting reliable information on the dynamic behaviour of real buildings. However it can be fastidious to link this information, i.e. eigenfrequencies and mode shapes, with a convenient representation that we can use to perform calculations. Many methods exists among which the continuous modelling that appears for instance in the seismic codes.
The advantage of the continuous modelling is mainly the simplicity of use. Furthermore, the most part of recent or not buildings presents a vertical periodicity that inclines them to such modelling, on condition that they have a sufficient number of stories (five at least).
The most common continuous beams are the pure bending beam and the pure shear beam. If some buildings can be associated to these models, the most part of them can not whether they are too or not enough slender, whether their inner structure combines bending elements and shear elements (like a beam connected to a frame). An other reason is the repartition of the observed natural frequencies: for a shear beam, this follows the arithmetic progression of odd integers (**1**$f_1$, $f_2$ = **3**$f_1$, $f_3$ = **5**$f_1$ …) and for a bending beam, the progression of square integers ($f_1$, $f_2$ = **(3/1.2)²** $f_1$, $f_3$ = **(5/1.2)²** $f_1$ …). For real buildings, this sequence is often intermediate between these two models. It is the reason why in this study, a Timoshenko beam representation is used.

[1] Laboratoire de Géophysique Interne et Tectonophysique, BP 53 - 38041 Grenoble Cedex 9 - France
Email: cmichel@obs.ujf-grenoble.fr
[2] DGCB/URA CNRS 1652, Ecole Nationale des Travaux Publics de l'Etat, Rue Maurice Audin, 69518 Vaulx-en-Velin, France
Email : stephane.hans@entpe.fr
[3] Laboratoire de Géophysique Interne et Tectonophysique, BP 53 - 38041 Grenoble Cedex 9 - France
Email: pgueg@obs.ujf-grenoble.fr
[4] DGCB/URA CNRS 1652, Ecole Nationale des Travaux Publics de l'Etat, Rue Maurice Audin, 69518 Vaulx-en-Velin, France
Email : claude.boutin@entpe.fr

In a first time, a brief theoretical of Timoshenko beam is presented. Then a case study follows, from full-scale experiment of three buildings of Grenoble city.

### 1.1 Theoretical description

In the framework of modal analysis, the different variables are expressed on the form A(x, t) = A(x) $e^{i\omega t}$, where x is the position along the beam, t the time and ω the angular frequency. In a Timoshenko beam, the transversal deformations result from two contributions :
   (i) The bending motion characterized by the section rotation α(x) , the bending stiffness EI of a section and the associated momentum M(x) = -EI α'(x),
   (ii) The shear motion characterized by the deflection β(x) (i.e. the angle between neutral axis and the section), the shear stiffness K of a section and the associated shear force T(x) = - K β(x).

The horizontal translation motion of the section, U, is related to the kinematical variables α and β by:

$$\alpha(x) = U'(x) + \beta(x) \tag{1}$$

Assuming the section rotation inertia J $\alpha^{(2)}(x)$ is negligible for a real building and denoting Λ the linear mass, the balance equations of a beam section are given by:

T'(x) = - Λω² U(x)
M'(x) = T(x)

The combination of these different relations lead to a Timoshenko beam description where translation motion U(x) is governed by the differential equation :

$$EI\ U^{(4)}(x) + (EI/K)\ \Lambda\omega^2\ U^{(2)}(x) = \Lambda\omega^2\ U(x) \tag{2}$$

### 1.2 Modal characteristics

Introducing the wavelength L = 2H/π (i.e. the first modal wavelength of a clamped-free shear beam), the general harmonic solution of (2) can be expressed on the form:

$$U(x) = a\cos(\delta_1 x/L) + b\sin(\delta_1 x/L) + c\cosh(\delta_2 x/L) + d\sinh(\delta_2 x/L) \tag{3}$$

where $\delta_1$ and $\delta_2$ are the positive numbers defined by :

$\delta_1^2 \delta_2^2 = \Lambda\omega^2 L^4 / EI$
$\delta_1^2 - \delta_2^2 = \Lambda\omega^2 L^2 / K$

In order to characterize the nature of the Timoshenko beam, the dimensionless parameter C, defined above with the structural characteristic of the section, is introduced:

$$C = EI / KL^2 \tag{4}$$

This parameter weights the importance of shear effect compared to the bending effect. Also the beam degenerates into a usual Euler-Bernoulli beam when C = 0 and to a pure shear beam when C = +∞. Obviously the height of the structure is taken into account in this evaluation.

The two dimensionless wave numbers $\delta_1$ and $\delta_2$ are also related by:

$$\frac{\delta_1^2 - \delta_2^2}{\delta_1^2 \delta_2^2} = C \quad \text{which implies:} \quad \delta_2 = \frac{\delta_1}{\sqrt{1 + C\delta_1^2}} \tag{5}$$

The condition of existence of non trivial solutions respecting the clamped-free boundary conditions i.e., at the base [U(0) = 0; α(0) = 0], and at the top [M(H) = 0; T(H) = 0], leads to the following wave number equation involving the unique parameter C:

$$2\delta_1^2\delta_2^2 + \delta_1\delta_2(\delta_1^2 - \delta_2^2)\sin(\delta_1\pi/2)\sinh(\delta_2\pi/2) + (\delta_1^4 + \delta_2^4)\cos(\delta_1\pi/2)\cosh(\delta_2\pi/2) = 0 \tag{6}$$

The infinite discrete series of roots $\delta_{1k}$ of (6) can be determined numerically for a fixed value of C. From the properties of trigonometric and hyperbolic functions, whatever the value of C, the series of roots $\delta_{1k}$ are close to the odd integers sequence 2k-1 (indeed odd integers are exact solutions for pure shear beam (C = +∞) and, in other cases, provide excellent approximations from the third mode). Finally the $k^{th}$ eigenfrequency is given by:

$$f_k = \frac{1}{2\pi L} \frac{\delta_{1k}^2}{\sqrt{\frac{\Lambda L^?}{EI} + \delta_{1k}^2 \frac{\Lambda}{K}}} \quad \text{where } \delta_{1k} \approx 2k-1 \text{ for } k \geq 2 \tag{7}$$

### 1.3 Relation between the parameter C and the eigenfrequencies sequence

A very interesting result is obtained when studying the eigenfrequencies sequence in function of C value. A bijective relation exists between C and the ratios $f_k/f_1$ (figure 1), it means :
  (i) for a given value of C corresponds only one sequence of $f_k/f_1$,
  (ii) conversely, for one ratio fk/f1 corresponds only one value of C.

This property is very useful, because the measurement of the two first eigenfrequencies is enough to determine the value of C, and thus the more efficient model of Timoshenko beam to describe the dynamics of the studied building.

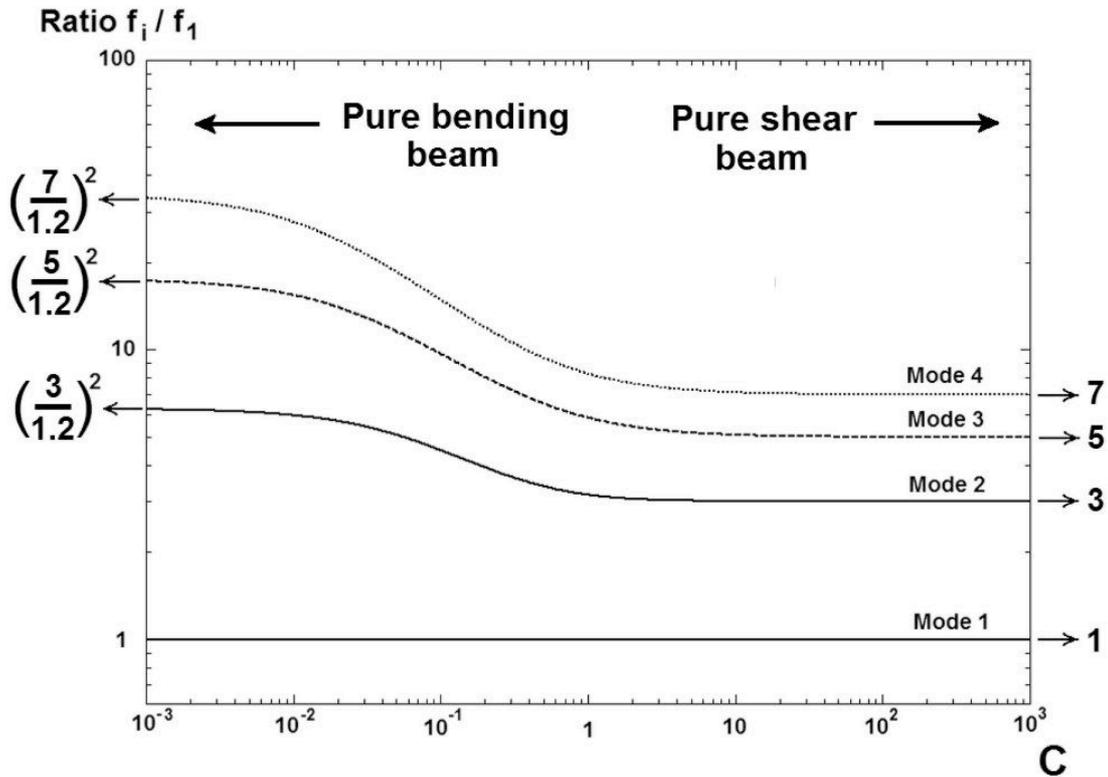

**Figure 1: ratio $f_k/f_1$ according to the value of C**

## 2. CASE-STUDY: THREE BUILDINGS OF GRENOBLE

### 2.1 The experimental part

#### 2.1.1 The tested buildings and the experimental procedure

Grenoble (France) is a dynamic city that increased in population during the late 60s. Many high-rise buildings were built at that time. The three Ile Verte tower are 28-story RC-buildings. Built between 1963 and 1967, these 100 m towers were the highest in Europe at that time. In March 2003, ambient vibrations were recorded at 15 different points in the Mont-Blanc tower (the central tower, Fig. 2) with a Cityshark II [Châtelain et al., 2006] and 6 velocimeters Lennartz 3D 5s. For this purpose, 3 datasets of 15 minutes at a frequency rate of 200 Hz were picked up. One reference sensor was left at the top floor whereas the others were roving along the stories in the staircase.
We followed the same process in December 2002 in one of the two ARPEJ towers, 15-story RC-buildings for students in the campus. In June 2005, we lead a full-scale experiment with the same recording material in the City Hall building [Michel and Guéguen, 2006], a 13-story RC building built in 1967.

We used the Frequency Domain Decomposition method (FDD) [Brincker et al., 2001] to analyse the modes of the structure. That means we estimated the Power Spectral Density (PSD) matrices, i.e. the Fourier transforms of the correlation matrices, and performed a singular value decomposition at each frequency. One can show that if the input is white noise, if the close modes are geometrically orthogonal and the damping is low, the structural frequencies are the peaks in the first singular value and the modal shapes are exactly the corresponding singular vectors.

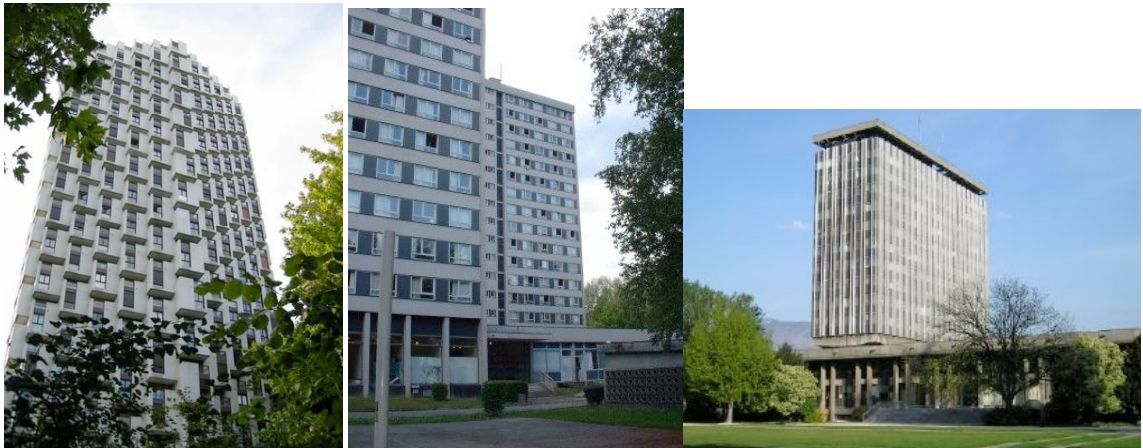

**Figure 2: The Mont-Blanc Tower, the ARPEJ Tower and the City Hall**

#### 2.1.2 Experimental results

For the Mont-Blanc and the ARPEJ towers, the sensor array did not allow determining the torsion modes. For the Mont-Blanc tower, we found the 6 first transverse modes and the 3 first longitudinal modes. The transverse modes can be found at 0.66, 2.69, 5.91, 9.11, 13.60 and 17.50 Hz and longitudinal modes at 0.85, 3.25 and 7.64 Hz. In the ARPEJ case, the 3 first modes in each direction were found at 1.17, 4.57, 8.52 Hz and 1.32, 5.03, 10.33 Hz in the transverse and longitudinal directions, respectively.
Thanks to the 35 measurement points in the City Hall, a reliable determination of the vibration modes including torsion modes was possible but we kept only the bending modes and 1D modal shapes for this study. Only the two first modes were excited enough to be determined at 1.22, 5.69 Hz and 1.15, 4.52 Hz in the transverse and longitudinal directions, respectively.

### 2.2 The theoretical interpretation

To determine the most adapted Timoshenko beam modelling, we used only the ratio $f_2/f_1$ between the two first experimental eigenfrequencies, because the measurement errors are a priori the smallest for the first modes. Thus only one value of the C-parameter is determined for each direction of vibration of the studied buildings. Results are showed in Table 1. With a C-parameter close to 0.2 (except for the City Hall transverse), the behaviour of all these buildings cannot be described only with a shear beam (C>5) or a cantilever beam (C<0.05). These first results confirm the Timoshenko beam model is more reliable for this kind of buildings.

In Table 2, the frequency sequence of the Timoshenko beam is compared to the experimental frequency sequence. Obviously, the two first theoretical eigenfrequencies are identical to the experimental ones in all cases, as a consequence of the determination method. A relatively good agreement is observed for the higher frequencies, even if the differences grow up with the mode number.

**Table 1: First frequency ratios and corresponding C-parameters for the three study-buildings**

| Building | Mont-Blanc Tower | | ARPEJ Tower | | City Hall | |
|---|---|---|---|---|---|---|
| Direction | Transverse | Longitudinal | Transverse | Longitudinal | Transverse | Longitudinal |
| Ratio $f_2/f_1$ | 4.07 | 3.80 | 3.90 | 3.81 | 4.66 | 3.93 |
| C | 0.167 | 0.239 | 0.209 | 0.234 | 0.083 | 0.200 |

**Table 2: Frequency sequence of the Timoshenko beam compared to the experimental values**

| | Direction | Frequency (Hz) | $f_1$ | $f_2$ | $f_3$ | $f_4$ | $f_5$ | $f_6$ |
|---|---|---|---|---|---|---|---|---|
| Mont Blanc Tower (28 stories) | Trans. | Experimental | **0.66** | **2.69** | 5.91 | 9.11 | 13.60 | 17.50 |
| | | Theoretical | 0.66 | 2.69 | 5.57 | 8.34 | 11.11 | 13.80 |
| | Long. | Experimental | **0.85** | **3.25** | 7.64 | - | - | - |
| | | Theoretical | 0.85 | 3.25 | 6.58 | - | - | - |
| ARPEJ (15 stories) | Trans. | Experimental | **1.17** | **4.57** | 8.52 | - | - | - |
| | | Theoretical | 1.17 | 4.57 | 9.33 | - | - | - |
| | Long. | Experimental | **1.32** | **5.03** | 10.33 | - | - | - |
| | | Theoretical | 1.32 | 5.03 | 10.20 | - | - | - |

The Figures 3 and 4 shows the experimental modal shapes compared to the Timoshenko beam modal shape with the experimental C-parameters. And in Table 3, the Modal Assurance Criterion (MAC) [Allemang and Brown, 1982] is calculated for each couple of experimental and corresponding theoretical mode shapes. This criterion quantifies the fit between these latest. So a very good agreement is observed for the three first modes in all cases, and a gap more and more important for the higher modes.

## 2.3 Comments

The used procedure leads to a representation agreeing well with the reality, for the first modes of vibration. The observed differences could be explained as:

(i) the not complete decoupling of the different directions; however, the inner structure of each building (not presented here) is very symmetrical and a coupling do not seem to be the real origin of the gaps;
(ii) the existence of a soil-structure interaction; it could be possible, because the measurements take into account the real vibrations of the buildings on their ground foundation; but in this case, the theoretical modal frequencies should be higher than the experimental ones, and, save for the transverse direction of ARPEJ, it is not what we can see.
(iii) Timoshenko model appears stiffer than the real system for the higher modes, so that an intern flexibility should exist and is not taken into account.
(iv) the scale separation is less and less respected for the higher modes, so the continuous representation is less and less justified for these modes (fourth and so on …)

The real reason is not well identified. A combination of previous explanations would be perhaps closer of the reality. Nevertheless, the representation by Timoshenko beam remains very efficient for the first modes of vibrations, the most important from a seismic point of view.

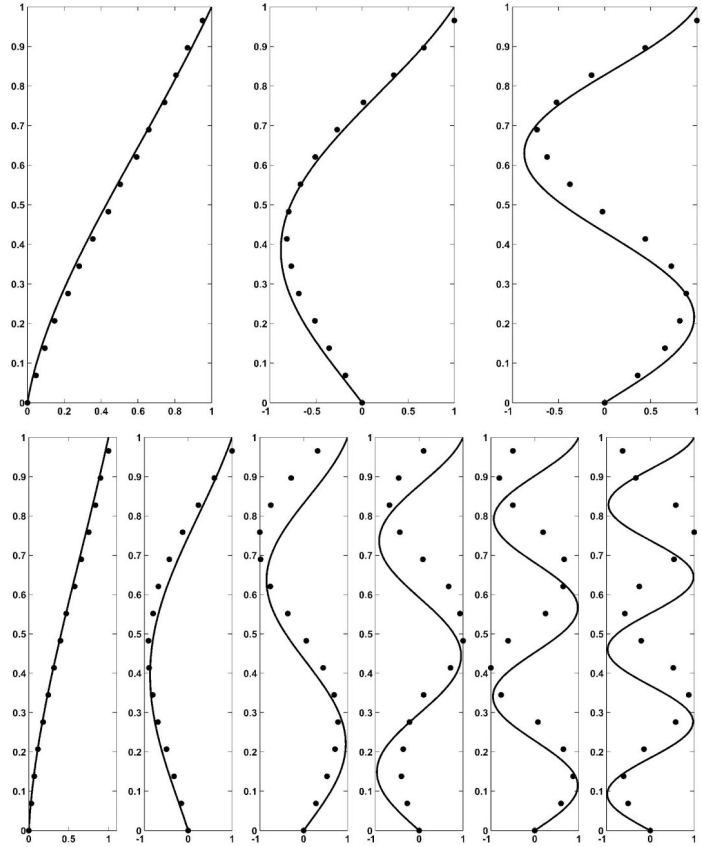

**Figure 3:** Experimental modal shapes compared to the Timoshenko beam model
with the experimental C-parameters. Mont-Blanc Tower - transverse and longitudinal directions

## 3.  CONCLUDING REMARKS

This work shows that the beam model integrates in a very synthetic way the essential parameters of the dynamic behavior. It appears that Timoshenko beam is very efficient to model the first vibration modes (three at least) of periodic buildings. Let us underline these modes are the most excited ones in the case of a seismic event, due to the frequency range of earthquake (0.1 to 15 Hz) and the participation factors (decreasing in $1/(2k+1)$ with the number k of the mode). Thus in a seismic assessment point of view, the use of this simplified model can give reliable information for a first level of diagnosis [Hans et al., 2005] [Boutin et al, 2005].

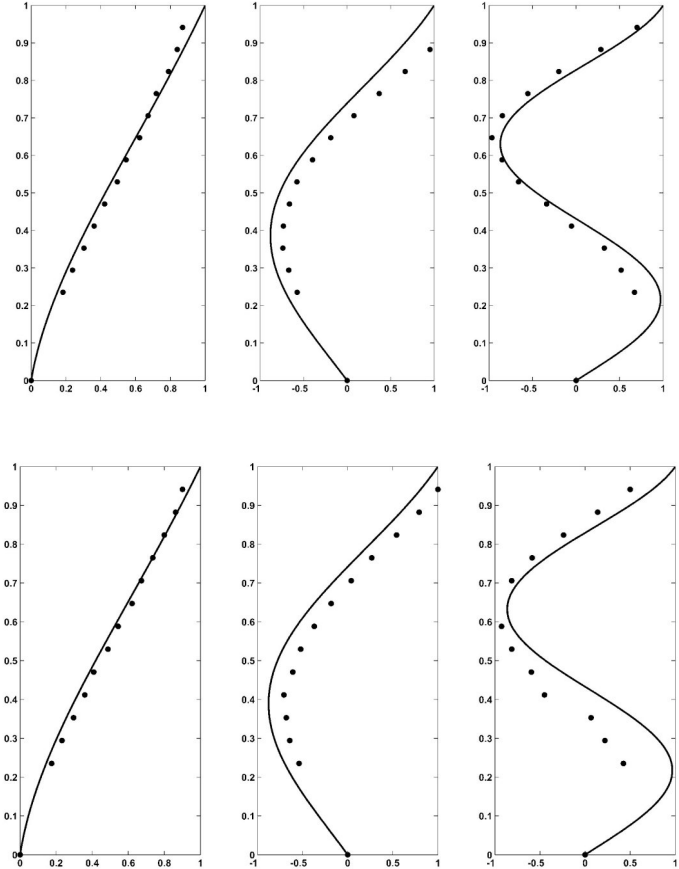

**Figure 4:** Experimental modal shapes compared to the Timoshenko beam model with the experimental C-parameters. ARPEJ - transverse and longitudinal directions.

| MAC value | | Mode 1 | Mode 2 | Mode 3 | Mode 4 | Mode 5 | Mode 6 |
|---|---|---|---|---|---|---|---|
| Mont-Blanc Tower | Longitudinal | 99.9% | 98.7% | 91.7% | - | - | - |
| | Transverse | 100.0% | 96.8% | 66.0% | 40.9% | 23.6% | 0.8% |
| ARPEJ Tower | Longitudinal | 99.8% | 92.0% | 70.1% | - | - | - |
| | Transverse | 99.7% | 88.6% | 93.5% | - | - | - |

**Table 3: Modal Assurance Criterion (MAC) between the experimental and Timoshenko modal shapes.**